\documentclass[journal=jpcl,manuscript=article]{achemso}
\usepackage{graphicx}
\usepackage{latexsym}
\usepackage{amsmath}
\usepackage{amssymb}
\usepackage{amsthm}
\usepackage{amsfonts}
\usepackage{color}
\usepackage{xcolor}
\usepackage{verbatim}
\usepackage{url}
\usepackage{array}
\usepackage{multirow}
\usepackage{courier}

\def\br{\mathbf{r}}
\def\icomp{\mathrm{i}}

\newcommand\refeq[1]{Eq.~(\ref{#1})}

\newcommand\marm[1]{\textcolor{black}{{#1}}}

\newcommand\onlinecite[1]{\hspace{-1 ex} \nocite{#1}\citenum{#1}} 

\graphicspath{{./Figures/}{./}}

\author{Pau Besal\'u-Sala}
\affiliation{Department of Chemistry and Pharmaceutical Sciences, Amsterdam Institute for Molecular and Life Sciences (AIMMS), Vrije Universiteit Amsterdam, De Boelelaan 1083, 1081 HV Amsterdam, The Netherlands}
\alsoaffiliation{Institut de Química Computacional i Catàlisi and Departament de Química, Universitat de Girona, 17003 Girona, Spain}

\author{Fabien Bruneval}
\affiliation{Universit\'e Paris-Saclay, CEA, Service de recherche en Corrosion et Comportement des Matériaux, SRMP, 91191 Gif-sur-Yvette, France}

\author{\'Angel Jos\'e P\'erez-Jim\'enez}
\affiliation{Department of Physical Chemistry, University of Alicante, E-03080 Alicante, Spain}

\author{\\Juan Carlos Sancho-Garc\'ia}
\affiliation{Department of Physical Chemistry, University of Alicante, E-03080 Alicante, Spain}

\author{Mauricio Rodr{\'i}guez-Mayorga}
\affiliation{Department of Physical Chemistry, University of Alicante, E-03080 Alicante, Spain}
\email{marm3.14@gmail.com}

\title{RPA, an accurate and fast method for\\ the computation of static non-linear optical properties}

\begin{document}


\begin{abstract}
The accurate computation of static non-linear optical properties (SNLOPs) in large polymers requires accounting for electronic correlation effects with a reasonable computational cost. The Random Phase Approximation (RPA) used in the adiabatic connection fluctuation theorem is known to be a reliable and cost-effective method to render electronic correlation effects when combined with the density-fitting techniques and the integration over imaginary frequencies. We explore the ability of the RPA energy expression to predict SNLOPs by evaluating RPA electronic energies in the presence of finite electric fields to obtain (using the finite difference method) static polarizabilities and hyper-polarizabilities. We show that RPA based on hybrid functional self-consistent field calculations yields as accurate SNLOPs as the best-tuned double-hybrid functionals developed today, with the additional advantage that RPA avoids any system-specific adjustment.
\end{abstract}


\maketitle


\section{Introduction}
The development of lasers in the 1960s led to the blossoming of non-linear optics (NLO). Nowadays, many technologies are based on non-linear optical effects triggered by the non-linear optical properties (NLOPs) of materials. For example, NLOPs have applications in optical signal processing~\cite{astill1991material}, ultra-fast switches~\cite{quemard2001chalcogenide}, sensors~\cite{gounden2020recent}, and laser amplifiers~\cite{mace1987aspects}, among others~\cite{castet2022predicting}. From the computational perspective, the prediction of static NLOPs (SNLOPs) for materials requires the evaluation of (hyper)-polarizabilities produced with accurate methods~\cite{de2014evaluation,champagne2018quantum,naim2023accelerated} able to describe the response of the many-body wavefunction $\Psi$ and in particular of the electronic density, 
$
n(\br) = N \int d\br _2 \ldots d\br _N \Psi ^* (\br, \br _2, \ldots ,\br _N) \Psi  (\br, \br _2, \ldots,\br _N),     
$ to the external fields produced by the incident light beams (see for example Ref.~\onlinecite{zhenyu2023}). The classic definition of the molecular static electric (hyper)-polarizabilities comes from the Taylor series expansion of the field-dependent dipole moment~\cite{chopra1989ab,buckingham1967quarterly,orr1971perturbation,ditchfield1970comparison} ${\boldsymbol \mu} ({\bf F})$ or energy~\cite{nakano1995size} $E({\bf F})$. In the case of the $E({\bf F})$, its components read as 
\begin{equation}
E ({\bf F}) = E({\bf 0})  - \sum _{i}  \mu _{i}  F_i - \frac{1}{2!}\sum _{ij}  \alpha _{ij}  F_i F_j - \frac{1}{3!} \sum _{ijk}  \beta _{ijk}  F_i F_j F_k -\frac{1}{4!} \sum _{ijkl}  \gamma _{ijkl}  F_i F_j F_k F_l + \dots   
\label{eq:dip_pol}
\end{equation}
with $i,j,k$ and $l$ being any Cartesian component (i.e. $x,y$ or $z$) and the expansion coefficients,
\begin{align}
\mu _{i} = - \frac{\partial  E }{\partial F_i } \Big|_{{\bf F} = {\bf 0}} \; , \; \alpha _{ij} &= - \frac{\partial ^2 E }{\partial F_i \partial F_j } \Big|_{{\bf F} = {\bf 0}} \; , \;   \beta _{ijk} = - \frac{\partial ^3 E}{\partial F_i \partial F_j  \partial F_k}  \Big|_{{\bf F} = {\bf 0}} \;,  \nonumber \\ \; \textrm{and} \; \; \gamma _{ijkl} &= - \frac{\partial ^4 E}{\partial F_i \partial F_j  \partial F_k \partial F_l} \Big|_{{\bf F} = {\bf 0}},  
\end{align}
being the dipole moment, the static polarizability, the first hyper-polarizability, and the second hyper-polarizability, respectively.
The (hyper)-polarizabilities can be alternatively obtained directly from the derivatives of the electronic density (see Ref.~\onlinecite{besalu2020new} and references therein). 

Within density functional theory (DFT), the electronic energy is a functional of the electronic density $E=E\left[ n(\br )\right]$~\cite{hohenberg_pr1964}. In practical applications using DFT, it is common to employ the Kohn-Sham (KS) scheme~\cite{kohn_pr1965}, which is based on expressing the electronic many-electron wavefunction as a single-determinant $\Phi$
and the density as $n(\br) = \sum ^{occ} _{i} |\phi _i ({\bf r})|^2$ where $\{\phi_i\}$ are the one-electron wavefunctions that solve the Kohn-Sham Hamiltonian (i.e. $\widehat{H}^{KS} \phi _i= \varepsilon _{i} \phi _i$). Unfortunately, within KS DFT the unknown exchange-correlation energy functional ($E_{xc}$) needs to be approximated. Numerous $E_{xc}$ functional approximations proposed in the literature~\cite{burke2013dft} also depend explicitly on the $\{\phi_i\}$ and $\{\varepsilon_i\}$, like the so-called (double)-hybrid functional approximations. In consequence, the total energy for the (double)-hybrid functional approximations including the contributions from the nuclei within the Born-Oppenheimer approximation~\cite{combes1981born} reads
\begin{align}
\label{eq:Etot}
 E \left[ \phi,\varepsilon \right] &= -\frac{1}{2}\sum ^{occ} _i \int d \br \; \phi ^* _i ({\bf r}) \nabla ^2 _\br \phi  _i ({\bf r}) + \int d\br \; v_\textrm{ext} (\br ) n(\br ) + \frac{1}{2} \int d\br \; d\br' \; \frac{n(\br )n(\br ')}{|\br - \br '|} + E_{nn-nf} \nonumber \\
 & + a E_x ^{\textrm{EXX}} \left[ \{ \phi \} \right] + (1-a) E_x ^{\textrm{DFT}} \left[n,\nabla n\right]+ b E_c ^{\textrm{PT}} \left[ \{ \phi ,\varepsilon \} \right] + (1-b) E_c ^{\textrm{DFT}} \left[n,\nabla n \right] ,
\end{align}
with $a$ and $b$ being the hybridization parameters, $E_{nn-nf}$ being the nucleus-nucleus interaction plus any nucleus-external field interaction, $v_\textrm{ext} (\br )$ being the time-independent external potential produced by the (bare) nuclei plus any electric field applied ${\bf F}$.
$E_x ^{\textrm{DFT}}$ ($E_c ^{\textrm{DFT}}$) is a DFT functional approximation that accounts for the exchange (resp. correlation) often depending also on the gradient of the electronic density $\nabla n$~\cite{perdew_prl1996}, $E_x ^{\textrm{EXX}}$ is the exact exchange obtained using $\Phi$ (i.e. the occupied orbitals), and $E_c ^{\textrm{PT}}$ is a term approximated from perturbation theory~\cite{gorling_prb1996} (i.e. $E_c ^{\textrm{PT}} \sim E_c ^{\textrm{MP2}}$) that accounts for electronic correlation effects~\bibnote{With MP2 being the second-order perturbation theory correction as proposer by Møller and Plesset~\cite{moller_pr1934}}. Unfortunately, in general, KS DFT functional approximations fail to predict accurate SNLOPs due to the delocalization error~\cite{bryenton2023delocalization}, the self-interaction error~\cite{bao2018self}, and the incorrect decay of the exchange-correlation kernels~\cite{choluj2018benchmarking}. By partially correcting these issues, the so-called range-separated functionals~\cite{stoll1985density,leininger1997combining,savin1996recent} such as LC-BLYP~\cite{iikura2001long}, CAM-B3LYP~\cite{yanai_cpl2004}, among others have shown to improve in the description of SNLOPs w.r.t. to their non-range separated counterparts. Further optimal tuning (OT) of range-separated functionals can yield better SNLOP predictions. Among all the (OT)-range-separated functionals, the T$\alpha$-LC-BLYP functional approximation, proposed by one of us,~\cite{besalu2020new}$^,$\bibnote{Built using the most popular OT technique that consists of enforcing the analogous of Koopmans' theorem for DFT~\cite{koopmans1934zuordnung,janak_prb1978}.} has proven to be the best approximation among the tested functionals for computing SNLOPs for a set of different organic and inorganic molecules. Stepping up to the highest rung that is currently in use in Jacob's ladder picture,~\cite{goerigk2014double,bremond2016nonempirical} we find the KS DFT functional approximations with a certain amount of MP2 correlation energy (a.k.a. double-hybrid functionals in the literature~\cite{grimme_jcc2006,bremond2011seeking,bremond2014communication}) such as B2PLYP~\cite{grimme_jcc2006}, PBE0-DH~\cite{bremond2011seeking}, and PBE-QIDH~\cite{bremond2014communication}, which have been proposed in an attempt to improve the non-local character of the correlation energy at the expense of increasing the computational cost. Let us stress that the ability of the most recently developed double-hybrid approximations to predict SNLOPs was still an open question as of today.

Another method to produce electronic energies consists in using the random phase approximation (RPA); setting $E_c ^\textrm{PT} \sim E_c ^\textrm{RPA}$ in~\refeq{eq:Etot} with the $E_c ^\textrm{RPA}$ given by the adiabatic connection fluctuation-dissipation theorem expression, which can be written using imaginary frequencies as~\cite{langreth_prb1977,furche2005fluctuation,fuchs_prb2002} 
\begin{equation}
E_c ^{\textrm{RPA}}[\chi _0] = \frac{1}{4 \pi} \int_{-\infty}^{+\infty} d \omega \,  \mathrm{Tr} \left\{ v\chi_0(\icomp \omega) + \mathrm{ln} \left[ 1 - v \chi_0(\icomp \omega) \right] \right\}.
\label{eq:rpa_phi}
\end{equation}
where $\textrm{Tr}$ indicates the trace, $v$ is the Coulomb interaction, and $\chi _0$ is the non-interacting (dynamic) polarizability that can be built in terms of the KS DFT $\{\phi_i\}$ and $\{\varepsilon_i\}$. RPA
has attracted much attention from the scientific community~\cite{dobson_prl1999,harl_prb2008,schimka_nm2010,lebegue_prl2010,ren_njp2012,ren_prb2009,bruneval_prl2012,bruneval_jctc2021,rodriguez2021coupling} due to its ability to account for the so-called non-dynamic electronic correlation effects~\cite{cohen2001dynamic} with a low computational cost (i.e. it scales as $M^3$ with $M$ being the size of the basis set~\cite{rojas_prl1995,foerster_jcp2011,kaltak_prb2020,wilhelm_jcpl2018,duchemin2019separable,duchemin2021cubic}). The usual procedure for computing RPA electronic energies involves performing a KS DFT calculation with a DFT functional approximation~\bibnote{That may include a certain amount of exact exchange.} and setting $b=0$ in \refeq{eq:Etot}. Once the self-consistent procedure is completed, the KS DFT $\{\phi_i\}$ and $\{\varepsilon_i\}$ are inserted in~\refeq{eq:Etot} with $a=1$ and $b=1$ (i.e. canceling all KS DFT contributions) to produce electronic energies. The good ratio of accuracy and computational cost of RPA makes it a very appealing approximation for the computation of SNLOPs for large systems:
it is the main focus of this work.


\section{Methodology}
\begin{figure}
    \centering
    \includegraphics[scale=0.6]{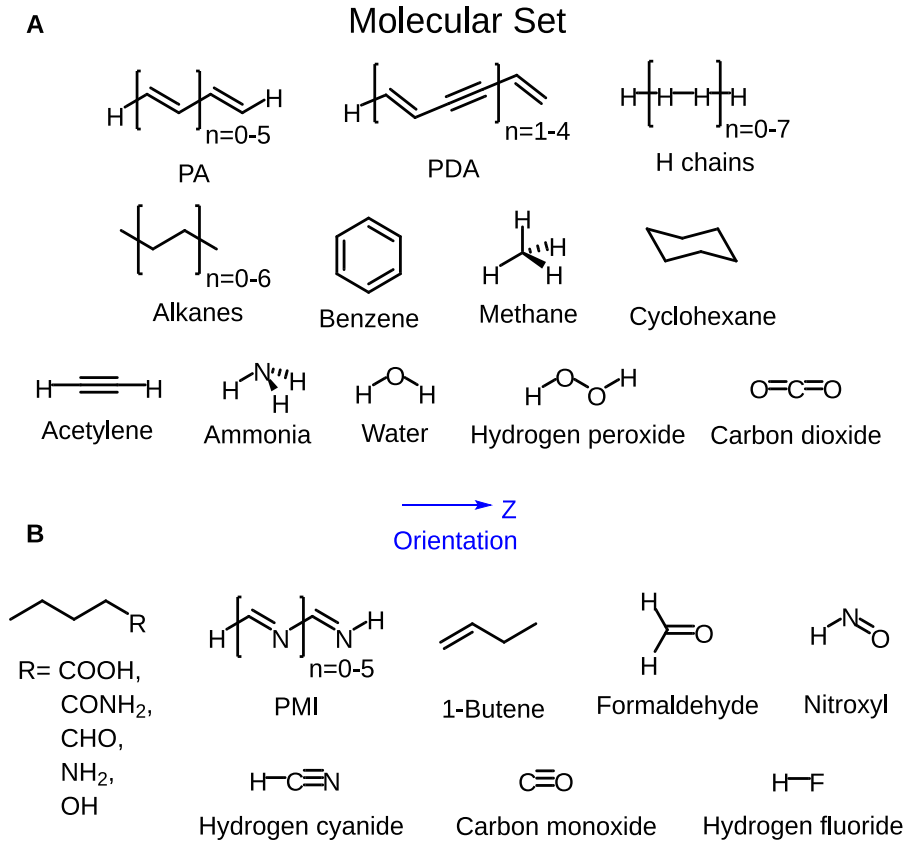}
    \caption{Chemical structures of the molecular set employed in this study including the orientation of the applied field. Subset A contains molecules that present their odd energy derivatives equal to 0 (i.e. $\mu _z = \beta _{zz}=0$) because they have inversion symmetry, while subset B is formed by molecules with all their energy derivatives different from 0. The geometries were taken from the supporting information of Ref.~\citenum{besalu2020new} (and are also available in Ref.~\citenum{molprolab}).}
    \label{fig:molecules}
\end{figure}

All implementations were performed in \texttt{MOLGW} code~\cite{bruneval_cpc2016,molgw_website}, where electric fields ${\bf F}$ are used for the computation of the numerical (hyper)-polarizabilities in~\refeq{eq:dip_pol} (see the Supporting Information for more details). The aug-cc-pVDZ basis set~\cite{dunning_jcp1989,woon_jcp1993,kendall_jcp1992} has been chosen to facilitate the comparison with previous studies~\cite{besalu2020new}. The set of systems employed in this work has been taken from Ref.~\onlinecite{besalu2020new}, where coupled-cluster singles and doubles with perturbative triples correction $[$CCSD(T)$]$ values were used as reference. In particular, 50 representative molecular
systems formed by atoms of the second period
and/or hydrogen (see Fig.~\ref{fig:molecules}) were chosen from Ref.~\onlinecite{besalu2020new}, including representative oligomers among the most typical $\pi$-conjugated NLO polymers, such as all-\textit{trans} polyacetylene (PA), all-\textit{trans} poly-methyneimine (PMI), and polydiacetylene (PDA). Lastly, some small organic and inorganic molecules, and hydrogen chains (known to be challenging systems for the calculation of SNLOPs) are also included. Interestingly, the studied set had been used for the construction of a long-range corrected range-separated hybrid, called T$\alpha$-LC-BLYP, that has been OT to compute the SNLOPs of the molecules of this set (see the Supporting Information for more details), allowing us to conduct a strict assessment of the RPA results. 

\section{Results}
The aim of this work is to scrutinize the performance of the RPA energy functional given in Eqs.~(\ref{eq:Etot}) and~(\ref{eq:rpa_phi}) to predict polarizabilities and hyper-polarizabilities and decide whether RPA can be used routinely in the computation of SNLOPs thanks to its (relatively) low computational cost~\cite{rojas_prl1995,foerster_jcp2011,kaltak_prb2020,wilhelm_jcpl2018}. Also, the double-hybrid functional approximations (i.e. B2PLYP, PBE0-DH, and PBE-QIDH) are considered to be the highest rung in Jacob's ladder picture; thus, it is time to appraise the performance of the most recently developed approximations for predicting SNLOPs. Hence, we have computed the $\mu_z$, $\alpha_{zz}$, $\beta_{zzz}$, and $\gamma _{zzzz}$ values for the systems introduced in the previous section using RPA, B2PLYP, PBE0-DH, and PBE-QIDH. In general, our results show that the absolute errors are dominated by $\gamma_{zzzz}$, where they present the largest deviations w.r.t. the CCSD(T) reference values.~\cite{besalu2020new} It is worth mentioning that the largest absolute errors for each property were obtained for the large systems like PDA4 (see the Supporting Information for more details) in line with the previous work~\cite{besalu2020new}.

Next, let us remark that RPA is computed in a non-self-consistent manner; hence, we must assess the role of the amount of $E_x ^{\textrm{EXX}}$ present at the underlying self-consistent KS DFT level when using RPA. In previous works~\cite{bruneval_jctc2019,denawi2023gw}, it has been shown that large values of $a$ in~\refeq{eq:Etot} tend to produce electronic energies that depend less on the starting point. Hence, we also aim to confirm if large $a$ values in~\refeq{eq:Etot} employed at the KS DFT starting point produce more accurate properties (i.e. SNLOPs). For this end,  we have taken as reference the PBEh($a$) hybrid functional~\cite{perdew1996rationale} and varied the amount of $E_x ^{\textrm{EXX}}$ (specifying this amount in parenthesis).

\begin{figure}[H]
    \centering
    \includegraphics[scale=0.4,angle=-90]{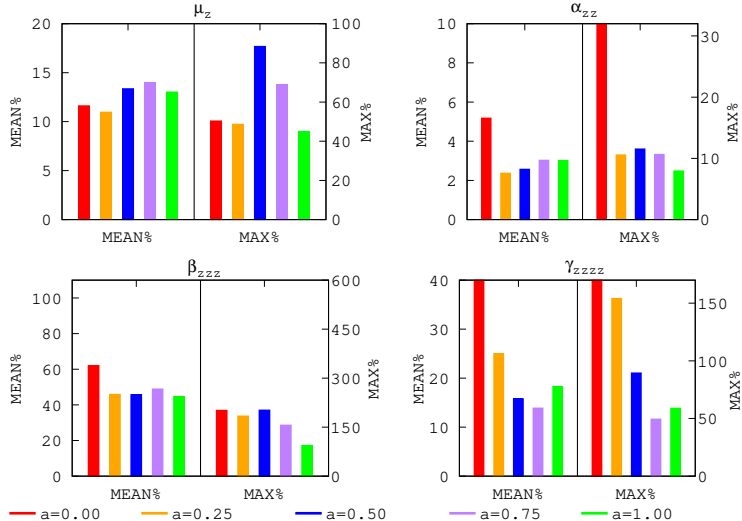}
    \caption{Mean relative errors (MEAN\%) and maximum relative errors (MAX\%)  obtained with RPA@PBEh($a$) for the predicted (hyper)-polarizabilities w.r.t. CCSD(T) for different values of the $a$ hybridization coefficient employed at the KS DFT level. Note: In the case of RPA@PBEh(0.0), the MAX\% for $\alpha_{zz}$ is 38.47\%, the MEAN\% of $\gamma_{zzzz}$ is 96.85\%, and the MAX\% of $\gamma_{zzzz}$ is 898.91\%. Thus, these values lie out of the scale.}
    \label{fig:err_hyb}
\end{figure}

In Fig.~\ref{fig:err_hyb}, we have collected MAX\% and MEAN\%~\bibnote{Notice that for RPA@PBEh(0.0), the MAX\% for $\alpha_{zz}$ is 38.47\%, the MEAN\% of $\gamma_{zzzz}$ is 96.85\%, and the MAX\% of $\gamma_{zzzz}$ is 898.91\%; hence, these values lie out of the scale used.} to be used as a yardstick for the predicted dipole moments and (hyper)-polarizabilities using RPA@PBEh($a$) for different $a$ values. Our results indicate that $\mu_z$ is better described using lower values of the $a$ hybridization coefficient at the KS DFT level. Nevertheless, $\mu_z$ is not strongly dependent on the amount of $E_x ^{\textrm{EXX}}$, as it is shown in Fig.~\ref{fig:err_hyb}. For $\alpha _{zz}$, the role of the amount of $E_x ^{\textrm{EXX}}$ employed at the KS DFT level becomes more relevant. Clearly, a small amount of exact exchange (such as $a=0.25$) employed at the KS DFT level for producing $\{\phi_i\}$ and $\{\varepsilon_i\}$ is able to drop the relative errors dramatically to approximately half of the values obtained when a non-hybrid functional is employed as starting point (recalling that  PBEh($0$) is the pristine PBE functional). In addition, notice that the $a=0.25$ value already produces MAX\% and MEAN\% that compare well with the same values obtained with larger amounts of $E_x ^{\textrm{EXX}}$ ($a> 0.25$); therefore, the role of the hybridization coefficient $a$ is reduced for $\alpha_{zz}$ as soon as some exact exchange is already present at the KS DFT level. Moving to the next derivative, $\beta_{zzz}$, it is remarkable that using $a=0.0$ (i.e. the pristine PBE functional) or $a=0.5$ as starting points yields similar MAX\% values. Although, we observe an important decrease (of approximately 50\%) in the MEAN\% using $a=1.0$, which indicates that large values of $a$ should be preferred for predicting $\beta_{zzz}$ accurately. In sharp contrast, the relative errors obtained for $\gamma_{zzzz}$ indicate that the role of the amount of $E_x ^{\textrm{EXX}}$ is crucial for its accurate description, where a small amount of exchange is fundamental to drop the errors by about 80\% w.r.t. the RPA@PBEh($0$) (i.e. RPA@PBE) ones. Our results show that an optimal value of $0.5<a<1.0$ can be chosen for dropping the relative errors when computing $\gamma_{zzzz}$ because the RPA@PBEh(0.5) and RPA@PBEh(1.0) errors are larger than the RPA@PBEh(0.75) ones. In general, our analysis of the dependence on the starting point reveals that some amount of $E_x ^{\textrm{EXX}}$ must be present at the KS DFT level to produce acceptable $\{\phi_i\}$ and $\{\varepsilon_i\}$ to enter the RPA energy expression. This is in line with the fact that DFT functionals with a large amount of $E_x ^{\textrm{EXX}}$ tend to be more accurate for SNLOP computations. The dependence on the starting point seems to be less evident than for electronic energies reported in our previous works~\cite{bruneval_jctc2019,denawi2023gw} because for each SNLOP an $a$ value can be selected leading to accurate values for that particular property, but not necessarily improving the description of the other ones, with $\mu_z$ and $\beta_{zzz}$ being less dependent on the starting point than $\alpha_{zz}$ and $\gamma_{zzzz}$. Lastly, we observe that large values of $a$ (i.e. $a>0.75$) seem to be able to lead to a compromise situation where $\mu_z$, $\alpha_{zz}$, $\beta_{zzz}$, and $\gamma_{zzzz}$ can all be reasonably well described with the same fixed $a$ value. 

From our previous analysis, we recognized that $a \in \left(0.5,1.0\right)$ could lead to an accurate prediction of $\beta_{zzz}$ and $\gamma_{zzzz}$ at the same time, the two properties that usually DFT tends to fail in larger proportion to evaluate. Indeed, we have found that a $a=0.85$ value leads to a good compromise situation for both properties, which allows us to introduce the RPA@PBEh(0.85) approximation for the description of SNLOPs. In Table~\ref{tbl:mae_re}, we have collected the mean absolute errors (MAE) in a.u., root-mean-square errors (RMSE) in a.u., maximum error (ME) in a.u., MEAN\%, and MAX\% values obtained with this method. As expected MAE, RMSE, and ME are dominated by large systems (e.g. PDA4) and their error increases rapidly with the order of the (numerical) energy derivative (i.e. they increase from $\mu_z$ to $\gamma_{zzzz}$). From Table~\ref{tbl:mae_re}, we observe that for the MAE, RMSE, and ME that the RPA@PBEh(0.85) approximation performs as well as the T$\alpha$-LC-BLYP functional for $\mu_z$, where the largest deviations are obtained for the PMI6 system. It is the best method for predicting $\alpha_{zz}$ with the best description of PDA systems than any other method and with the largest deviations obtained for PA systems. It deteriorates some for computing $\beta_{zzz}$, where the largest deviation is that of the PMI6 system. And, finally, it performs similarly to LC-BLYP functional for $\gamma_{zzzz}$ with the largest errors obtained for the PDA systems. Nevertheless, we must stress once more that for computing SNLOPs of molecules of different sizes, the most important errors are the relative errors (represented by MEAN\% and MAX\% in Table~\ref{tbl:mae_re}), where the RPA@PBEh(0.85) approximation performs better than any non-OT-range-separated functional (e.g. it is better than LC-BLYP) for dipole moments presenting the largest deviations for small systems (i.e. HNO and CO). Moving on to the computation of $\alpha_{zz}$, RPA@PBEh(0.85) is the best method among all the tested ones, including OT-range-separated hybrids. For all the functionals tested, the largest errors on $\alpha_{zz}$ were produced by PA and PDA systems (in agreement with the results obtained for the MAE, RMSE, and ME). Remarkably, despite PA/PDA causing the highest errors for RPA@PBEh(0.85), those errors are lower compared to other functionals, explaining the success of RPA@PBEh(0.85) on the polarizability calculations. For $\beta_{zzz}$, RPA@PBEh(0.85) can also be considered the best method because it shows the lowest maximum percentage error, but on the contrary to $\alpha_{zz}$, in this case, there is not a specific family of molecules causing the error but the overall. Finally, for $\gamma_{zzzz}$, RPA@PBEh(0.85) is the second-best method behind the (system adapted) T$\alpha$-LC-BLYP functional. This is expected as T$\alpha$-LC-BLYP was specifically designed to compute $\gamma_{zzzz}$ on this particular set of molecules~\cite{besalu2020new}. Interestingly, the largest deviations of RPA@PBEh(0.85) are obtained for small systems (i.e. CO and H$_2$O), while relative errors lower than 32\% for all the polymers studied were obtained, and with excellent performance for Hydrogen chains where relative errors are lower than 8\%, which praises its good performance for predicting $\gamma_{zzzz}$.  It is worth stressing again that RPA@PBEh(0.85) does not involve any extra re-optimization of parameters for each system, which makes it a very competitive approximation for computing SNLOPs when compared with the best KS DFT functional currently in use~\cite{besalu2020new}. 

\begin{table}[ht]
  \caption{Mean absolute errors (MAE) in a.u., root-mean-square errors (RMSE) in a.u., maximum error (ME) in a.u., mean relative errors (MEAN\%), and maximum relative errors (MAX\%) for the predicted (hyper)-polarizabilities w.r.t. CCSD(T) values.\textsuperscript{\emph{a}}}
  \label{tbl:mae_re}
  \begin{tabular}{llrrrr}
    \hline
    Property  & &  T$\alpha$-LC-BLYP\textsuperscript{\emph{a}} & LC-BLYP & PBE-QIDH & RPA@PBEh(0.85)  \\
    \hline
    \multirow{4}{*}{$\mu_{z}$} & MAE (a.u.) & 0.12 & 0.16 & 0.12 & 0.11 \\
     & RMSE (a.u.) & 0.20 & 0.25  & 0.20 & 0.19 \\
     & ME (a.u.) & 0.53 & 0.67 & 0.52 & 0.50 \\
     & MEAN\% & 5.03 & 29.89 & 13.88 & 13.67 \\
     & MAX\%  & 67.11 & 174.32 & 46.36  & 52.61 \\ \hline
    \multirow{4}{*}{$\alpha_{zz}$} & MAE (a.u.) & 9.61 & 7.16 & 12.32 & 5.34 \\
     & RMSE (a.u.) & 17.54 & 99.63 & 13.49 & 10.89 \\
     & ME (a.u.) & 63 & 48 & 126 & 41 \\
     & MEAN\% & 5.66  & 4.56 & 5.45 & 3.07 \\
     & MAX\%  & 13.15 & 16.19 & 20.38 & 9.73 \\ \hline
    \multirow{4}{*}{$\beta_{zzz}$} & MAE (a.u.) & 101.08 & 52.30 & 96.35 & 142.10 \\
     & RMSE (a.u.) & 276.54 & 382.42 & 140.09 & 348.27 \\
     & ME (a.u.) & 1.11$\times$10$^3$ & 1.50$\times$10$^3$ & 5.55$\times$10$^2$ & 1.24$\times$10$^3$ \\
     & MEAN\% & 73.59  & 61.14  & 39.20  & 49.34  \\
     & MAX\%  & 433.48 & 413.13 & 258.14 & 121.06 \\ \hline
    \multirow{4}{*}{$\gamma_{zzzz}$} & MAE (a.u.) & 1.88$\times$10$^4$ & 4.06$\times$10$^4$ & 1.51$\times$10$^5$ & 6.58$\times$10$^4$ \\
    & RMSE (a.u.) & 5.85$\times$10$^4$ & 9.22$\times$10$^4$ &  6.35$\times$10$^5$ & 2.60$\times$10$^5$ \\
     & ME (a.u.) & 3.22$\times$10$^5$ & 3.43$\times$10$^5$ & 4.16$\times$10$^6$ & 1.64$\times$10$^6$ \\
     & MEAN\% & 5.92   & 19.44  & 19.13  & 13.90 \\
     & MAX\%  & 29.32  & 50.73  &  78.45 & 49.50 \\
    \hline
  \end{tabular}

  \textsuperscript{\emph{a}} CCSD(T) and T$\alpha$-LC-BLYP values taken from Ref.~\onlinecite{besalu2020new}.
\end{table}

\begin{figure}[H]
    \centering
    \includegraphics[scale=0.4,angle=-90]{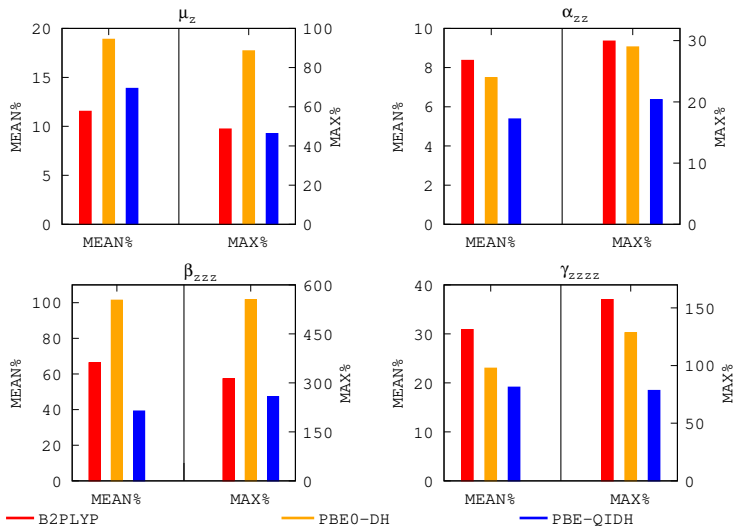}
    \caption{Mean relative errors (MEAN\%) and maximum relative errors (MAX\%) for the predicted (hyper)-polarizabilities w.r.t. CCSD(T) for B2PLYP, PBE0-DH, and PBE-QIDH.}
    \label{fig:err_dhyb}
\end{figure}

Among the double-hybrid functional approximations, the MAE, RMSE, and ME are again dominated by large systems (see the supporting information for more details). Nevertheless, focusing on relative errors we observe from Fig.~\ref{fig:err_dhyb} that PBE-QIDH outperforms B2PLYP and PBE0-DH for all polarizabilities and hyper-polarizabilities. Actually, only for predicting dipole moments the PBE0-DH functional is the best one (closely followed by PBE-QIDH). Interestingly, the PBE-QIDH functional presents the largest amount of $E_x ^{\textrm{EXX}}$, which confirms that the role of the exact exchange is fundamental for an accurate description of SNLOPs. Since PBE-QIDH is the best double-hybrid functional approximation employed in this work, we have collected in Table~\ref{tbl:mae_re} the MAE, RMSE, ME, MEAN\%, and the MAX\% obtained using this functional approximation. Clearly, focusing on relative errors, PBE-QIDH performs as well as RPA@PBEh(0.85) for $\mu_z$, it is very competitive for computing $\alpha_{zz}$ with errors that lie close to the T$\alpha$-LC-BLYP ones, it is a really good method for computing $\beta_{zzz}$ because its MEAN\% is the lowest one (39.2\%), and, finally, it compares well with the best (non-system adapted) range-separated functional (i.e. LC-BLYP) for predicting $\gamma_{zzzz}$. In general, PBE-QIDH results indicate that it is also a competitive method for computing SNLOPs, but involving an increase in the computational cost due to the evaluation of the MP2 correlation energy (that can rapidly increase with system size). Finally, let us mention that the increase in the complexity of the functional approximation due to the inclusion of a double hybrid scheme needs an adequate selection of the hybridization coefficients. In fact, the truly non-empirical design of PBE-QIDH (without any fitted parameter), as defined by some of us~\cite{bremond2016nonempirical,bremond2014communication}, leads to a large amount of exact exchange present in this functional ($a=3^{-{1/3}}\sim 0.69336$). Consequently, the large amount of $E_{x} ^{\textrm{EXX}}$ reduces fundamental errors (such as the delocalization error, the self-interaction error, among others) that can play an important role in the description of properties using KS DFT approximations as we show in this work.  

\begin{figure}[H]
    \centering
    \includegraphics[scale=0.4,angle=-90]{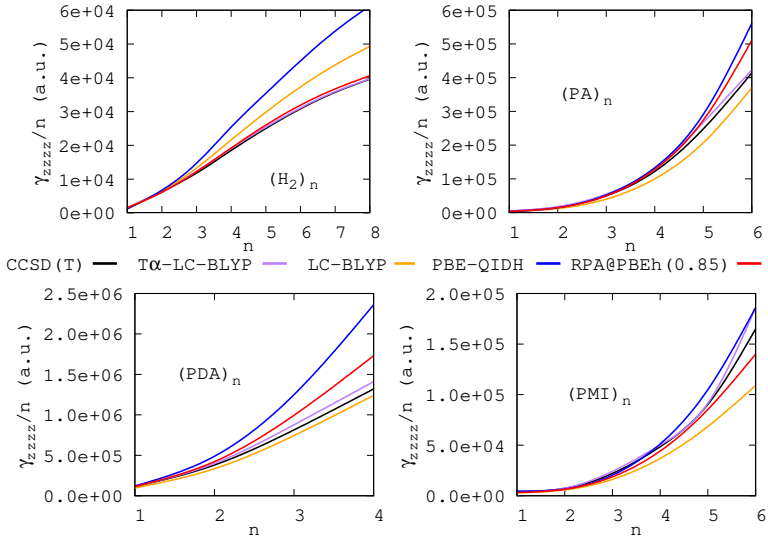}
    \caption{$\gamma_{zzzz}$ values per number of monomers against the number of monomers ($n$) for Hydrogen chains, PA, PDA, and PMI.}
    \label{fig:gamma_zzzz_n}
\end{figure}

To rationalize the results, let us recall that the functional expression used in this work for the RPA electronic energy employs (as it is usually done) 100\% of the exact exchange (i.e. $a=1$ in~\refeq{eq:Etot}); therefore, the well-known errors~\cite{bryenton2023delocalization,besalu2020new} that are responsible for the usual failures of KS DFT in the description of SNLOPs are drastically reduced. Moreover, the RPA is known to be equivalent to the direct ring-coupled-cluster doubles method~\cite{ren_prb2013}, which ensures a reasonable description of the so-called dynamic electronic correlation effects; approaching the coupled-cluster description for these effects. On the other hand, the results obtained with the double-hybrid functional approximations indicate that PBE-QIDH is the best double-hybrid KS DFT approximation for computing SNLOPs (among the ones tested in this work). The good performance of PBE-QIDH can be justified due to a large amount of exact exchange present in this functional ($a=3^{-1/3}\sim 0.69336$), while B2PLYP contains a lower amount of $E_x ^{\textrm{EXX}}$ ($a=0.53$) and PBE0-DH even a lower value ($a=0.5$). Indeed, the increase in the amount of exact exchange makes the PBE-QIDH reduce the delocalization error and the self-interaction error, which leads to a very competitive performance in the computation of polarizabilities and hyper-polarizabilities using this functional approximation. Finally, comparing PBE-QIDH performance for predicting SNLOPs with that of RPA@PBEh(0.85), see Table~\ref{tbl:mae_re}, the latter overtakes the former with less computational expense, which makes RPA@PBEh(0.85) a very attractive approximation for computing these properties. \marm{It is also interesting to gain further insights about the quality of the predicted $\gamma_{zzzz}$ when the number of monomers is increased (recalling that it is the most challenging property). To that end, we have plotted in Fig.~\ref{fig:gamma_zzzz_n} the $\gamma_{zzzz}/n$ values obtained with CCSD(T), T$\alpha$-LC-BLYP, LC-BLYP, PBE-QIDH, and RPA@PBEh(0.85) against the number of monomers ($n$) for polymeric systems. From Fig.~\ref{fig:gamma_zzzz_n} we notice that T$\alpha$-LC-BLYP and LC-BLYP perform better than RPA@PBEh(0.85) when $n$ increases for PA and PDA polymers when compared with the CCSD(T) reference values (with PBE-QIDH showing the largest deviations for these systems, despite being the best double-hybrid among the ones employed in this work). For the PMI polymers, the performance of RPA@PBEh(0.85), T$\alpha$-LC-BLYP, and PBE-QIDH is comparable when $n$ increases; with LC-BLYP showing the largest deviations for these polymers. Finally, in the case of the Hydrogen chains, we notice that RPA@PBEh(0.85) is a very accurate approximation to reproduce the CCSD(T) values being highly competitive with the T$\alpha$-LC-BLYP results. Overall, the RPA@PBEh(0.85) approximation is more stable than LC-BLYP and PBE-QIDH when the monomer is changed and $n$ increased; thus, it is only outperformed by T$\alpha$-LC-BLYP for the polymers studied. Similar results were obtained for $\alpha_{zz}/n$ when $n$ is increased, where the RPA@PBEh(0.85) approximation is shown to be the best approximation to reproduce the reference values (see the Supporting Information for more details).}

In summary, the RPA@PBEh(0.85) approximation is a simple, accurate, and fast method that can be used routinely in the computation of static non-linear optical properties. Its low computational cost~\cite{rojas_prl1995,foerster_jcp2011,kaltak_prb2020,wilhelm_jcpl2018,duchemin2019separable,duchemin2021cubic} \marm{(it scales as $M^3$, with $M$ being the size of the basis set, which is lower than the $M^5$ formal scaling of the double-hybrids or the $M^7$ scaling of the CCSD(T) method~\cite{yoshioka2021solving})} makes it a very competitive approximation outperforming KS DFT functionals approximations. We have also shown that PBE-QIDH functional performs well and is the best (global) double-hybrid KS DFT functional approximation (currently available) for computing SNLOPs, at the expense of increasing the computational cost. Since the PBE-QIDH functional approximation is a functional that does not involve a system-dependent optimization, its universality also facilitates its usage on different systems. \marm{Finally, let us comment, that the predictions of dynamic properties and how these findings could be transferable to those cases will remain open questions that need to be addressed in future work.} 


\begin{acknowledgement}
A. J. P.-J., J. C. S.-G., and M.R.-M. acknowledge the Ministerio de Ciencia e Innovación de Espa\~na for Grant No. PID2019-106114GB-I00. P. B.-S. acknowledges the financial support received from the Vrije Universiteit Amsterdam. F.B. and M.R.-M. acknowledge the financial support provided by the Cross-Disciplinary Program on Numerical Simulation of the French Alternative Energies and Atomic Energy Commission (CEA) (\textsc{ABIDM} project). We also acknowledge the computational resources of the Institut de Qu\'imica Computacional i Catàlisi of the Universitat de Girona. Finally, the authors thank {\'E}. Br\'emond for the discussions related to the implementation of double-hybrid KS DFT functionals.
\end{acknowledgement}


\begin{suppinfo}
Dipole moments, polarizabilities, and hyper-polarizabilities are given in the Supporting Information (as a spreadsheet). \marm{Also, in the Supporting Information we include additional computational details, information about the construction of the T$\alpha$-LC-BLYP approximation, the analysis of the performance of RPA@PBEh(0.85) in the prediction of $\alpha_{zz}/n$ for the polymeric systems, and further analysis of the predicted SNLOPs based on the size by magnitudes.}
\end{suppinfo}


\section*{Graphical abstract}
\begin{figure}[H]
    \centering
    \includegraphics[scale=0.03]{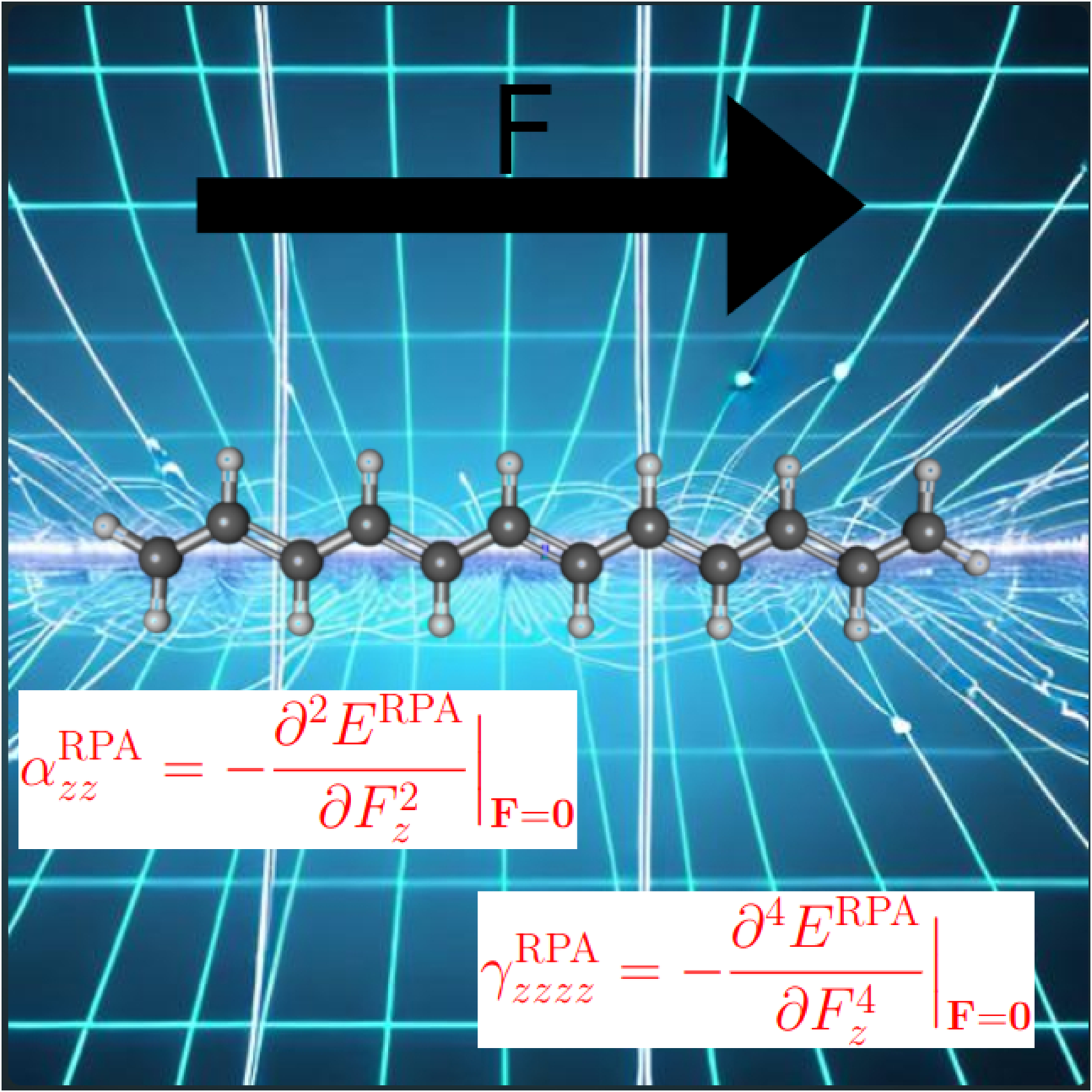}
    \caption{TOC}
    \label{fig:toc}
\end{figure}


\bibliography{rpa_pol.bib,nlops.bib}

\begin{mcitethebibliography}{69}
\bibitem[Not()]{Note-3}
That may include a certain amount of exact exchange.\relax
\mciteBstWouldAddEndPunctfalse
\mciteSetBstMidEndSepPunct{\mcitedefaultmidpunct}
{}{\mcitedefaultseppunct}\relax
\EndOfBibitem
\bibitem[Not()]{Note-4}
Notice that for RPA@PBEh(0.0), the MAX\% for $\alpha_{zz}$ is 38.47\%, the
  MEAN\% of $\gamma_{zzzz}$ is 96.85\%, and the MAX\% of $\gamma_{zzzz}$ is
  898.91\%; hence, these values lie out of the scale used.\relax
\mciteBstWouldAddEndPunctfalse
\mciteSetBstMidEndSepPunct{\mcitedefaultmidpunct}
{}{\mcitedefaultseppunct}\relax
\EndOfBibitem
\end{mcitethebibliography}


\end{document}